\documentclass[twocolumn,aps,floatfix,nofootinbib]{revtex4} 
\usepackage{amsmath}
\usepackage{ amssymb }
\usepackage{graphicx}
\usepackage{hyperref}
\hypersetup{colorlinks=true,linkcolor=blue,citecolor=red}

\begin{document}

\title{Universal Properties of Galactic Rotation Curves and a First Principles Derivation of the Tully-Fisher Relation}

\author{James~G.~O'Brien$^{1}$, Thomas~L.~Chiarelli$^{2}$, and Philip~D.~Mannheim$^{3}$}

\affiliation{$^{1}$Department of Sciences, Wentworth Institute of Technology, Boston, MA
02115  obrienj10@wit.edu\\
$^{2}$Department of Electromechanical Engineering, Wentworth Institute of Technology, Boston, MA
02115  chiarellit@wit.edu\\
$^{3}$Department of Physics, University of Connecticut, Storrs, CT 06269
 philip.mannheim@uconn.edu}

\begin{abstract}
In a recent paper McGaugh, Lelli, and Schombert showed  that in an empirical plot of the observed centripetal accelerations in spiral galaxies against those predicted by the Newtonian gravity of the luminous matter in those galaxies the data points occupied a remarkably narrow band. While one could summarize the mean properties of the band by drawing a single mean curve through it, by fitting the band with the illustrative conformal gravity theory with fits that fill out the width of the band we show here that the width of the band is just as physically significant. We show that at very low luminous Newtonian accelerations the plot can become independent of the luminous Newtonian contribution altogether, but still be non-trivial due to the contribution of matter outside of the galaxies (viz. the rest of the visible universe). We present a new empirical plot of the difference between the observed centripetal accelerations and the luminous Newtonian expectations as a function of distance from the centers of galaxies, and show that at distances greater than 10 kpc the plot also occupies a remarkably narrow band, one even close to constant. Using the conformal gravity theory we provide a first principles derivation of the empirical Tully-Fisher relation.

\end{abstract}

\date{October 3, 2017}

\maketitle

\section{Introduction} 

In a recent study McGaugh, Lelli, and Schombert  (MLS) \cite{McGaugh2016} presented an empirical plot of the observed centripetal accelerations ($g(OBS)$) of points in a wide class of spiral galaxies versus the luminous Newtonian expectations ($g(NEW)$) for those points. While the plot does not contain any information that is not already contained in plots of individual galactic rotation curves, the utility of the plot is that it allows one to include the data from every single galaxy in one and the same figure. The plot thus enables one to encapsulate a large amount of galactic rotation curve data in a single plot, doing so in a way that allows one to identify regularities in galactic rotation curve data that hold for all spiral galaxies. Inspection of the $g(OBS)$ versus $g(NEW)$ plot that we present in Fig. (\ref{gobsgbar}) immediately reveals three striking features. First, as noted by Milgrom in his development of the MOND theory \cite{Milgrom1983a}, the departure from $g(NEW)$ first sets in when $g(OBS)$ drops below a universal acceleration scale of order $10^{-10}$ m~s$^{-2}$. Second, when there are departures they occupy a remarkably small region in the plot. And third, as noted in \cite{McGaugh2016}, these departures would appear to be quite tightly correlated with the luminous Newtonian prediction. To quantify such a possible correlation, MLS made a one-parameter best mean fit to the plot in Fig. (\ref{gobsgbar}) with a fundamental acceleration parameter $g_0$, and found a good fit with the function  $g(OBS)=g(MLS)$ where
\begin{eqnarray}
 g(MLS)=\frac{g(NEW)}{[1-\exp(-(g(NEW)/g_0)^{1/2})]}, 
 \label{A1}
 \end{eqnarray}
and extracted a value $g_0=1.20\times 10^{-10}$ m s$^{-2}$. In this paper we shall evaluate the results of MLS and reach some new conclusions.

\section{The Data Analysis}

\begin{figure}[htpb!]
  \centering
  \includegraphics[width=3.2in,height=1.0in]{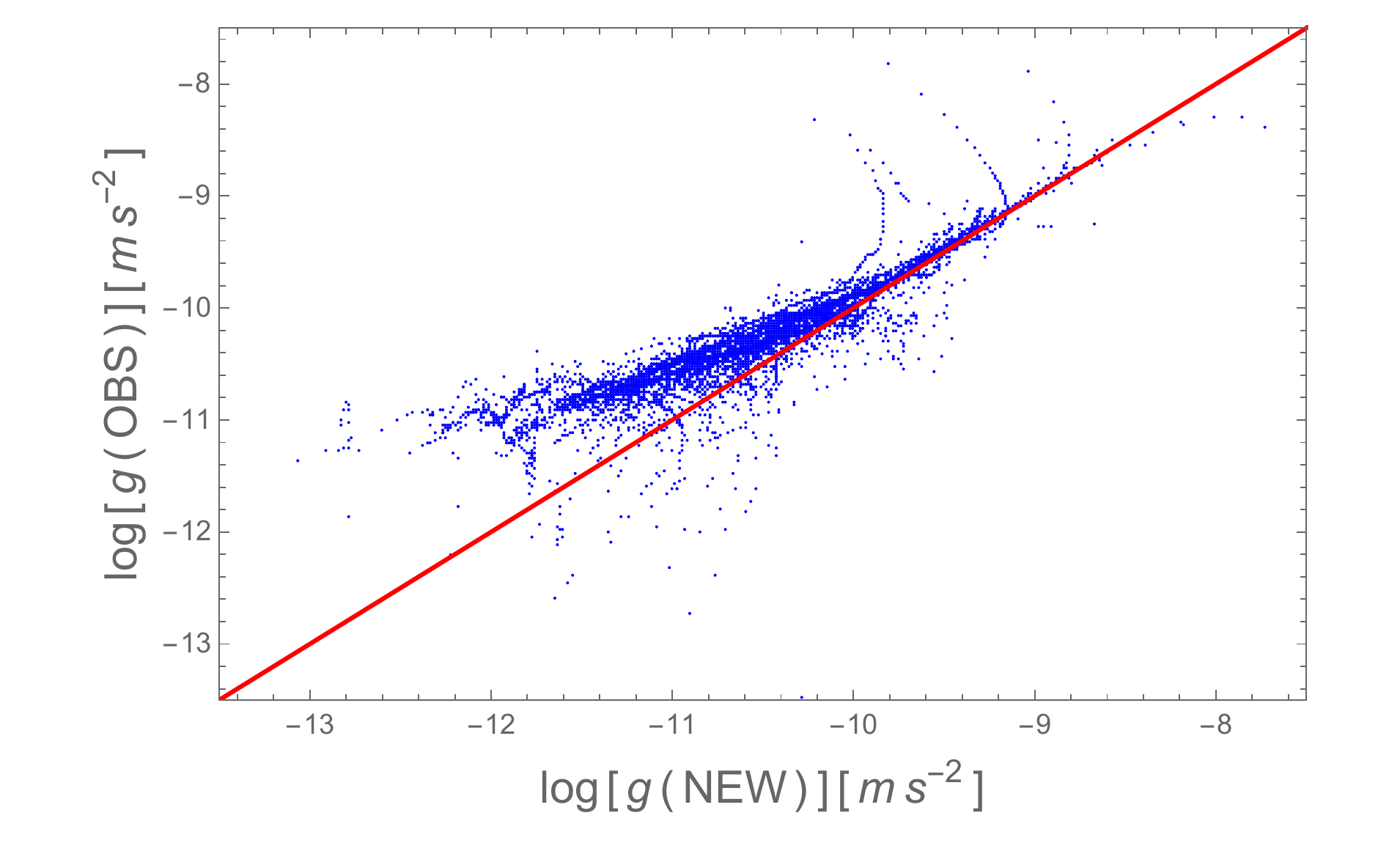}
  \caption{The $g(OBS)$ versus $g(NEW)$ plot and the $g(MLS)$  fit (dotted curve) to it. The solid diagonal is the line $g(OBS)=g(NEW)$.}
   \label{gobsgbar}
\end{figure}

In trying to produce a $g(OBS)$ versus $g(NEW)$ plot there are two key variables, the distance to each galaxy (needed to fix distances $R$ from galactic centers in both $g(OBS)=v^2_{OBS}/R$ and $g(NEW)$),  and the visible mass of each galaxy (needed for $g(NEW)$). While uncertainties in these quantities affect detailed fitting of both galactic rotation curves and Fig. (\ref{gobsgbar}), they do not affect the shapes of galactic rotation curves or the general structure of Fig. (\ref{gobsgbar}), with there always being departures from a $g(OBS)=g(NEW)$ curve. 

While the analysis we present follows MLS in the main, we use a somewhat larger sample of spiral galaxies and a somewhat different methodology to model the luminous matter  Newtonian contribution $g_{NEW}(R)$ to centripetal accelerations in galaxies. The original analysis of \cite{McGaugh2016} consisted of 153 spiral galaxies and a total of 2693 rotation curve data points. The sample we consider consists of 207 spiral galaxies, and a total of 5791 data points, many of which are also in the sample studied in \cite{McGaugh2016}. Our sample consists of the 141 galaxies that were studied in \cite{Mannheim2011} together with a set of 26 LITTLE THINGS galaxies \cite{Oh2015} and 40 galaxies from the Spitzer Photometry and Accurate Rotation Curves (SPARC) set studied in \cite{McGaugh2016}. In the following we will find it instructive to break the sample up into high surface brightness (HSB) and low surface brightness and dwarf galaxies (collectively LSB galaxies in the following since many LSBs are dwarfs), with our sample containing 56 HSB galaxies with 2870 points and 151 LSB galaxies with 2921 points. 

To model the luminous matter contribution we follow the procedure used in the successful fitting to individual rotation curves that was presented in \cite{Mannheim2011}. Specifically, we take the luminous matter optical disks to have surface brightness $\Sigma(R)=\Sigma_0\exp(-R/R_0)$ with scale length $R_0$. We take the HI gas in the galaxy to also be an exponential with scale length equal to four times the optical disk $R_0$, and multiply the HI gas mass by 1.4 to account for helium. We include bulges for the few galaxies in our sample that have them. We use reported optical disk scale lengths, and when scale lengths are reported in more than one filter in the main we take the longest wavelength available. For the distances to galaxies we use the mean values reported in the NASA/IPAC Extragalactic Database (NED) and allow for up to its stated one standard deviation variation in the few cases where it helped with the individual rotation curve fits. Since the study of \cite{McGaugh2016} leads to the possible presence of a universal acceleration scale in the data it is necessary to know the absolute distances to galaxies as well as possible, with our use of the NED data providing a uniform benchmark for these distances. For the inclinations of the galaxies we allowed up to the variation reported in observations, though this only affected a few of the galaxies. With each star putting out a Newtonian potential of the form $V_{{ NEW}}^*=-\beta^*c^2/r$ where $\beta^*=M_{\odot}G/c^2=1.48\times 10^5$ cm,  the contribution of an exponential disk with $N^*$ stars is of the form
\begin{eqnarray}
g_{{NEW}}(R)=\frac{N^*\beta^*c^2 R}{2R_0^3}[I_0\left(x
\right)K_0\left(x\right)-
I_1\left(x\right)
K_1\left(x)\right)],
\label{A2}
\end{eqnarray} 
where $x=R/2R_0$. Use of this formula in the 141 galaxy conformal gravity theory fits studied in \cite{Mannheim2011} and in conformal gravity fits to the individual rotation curves of the additional 66 galaxies in our sample (which we shall report on elsewhere) enabled us to extract out a value for $N^*$ (and thus a mass to light ratio $M/L$) for each galaxy. As well as being typical of the $M/L$ ratios ordinarily obtained in rotation curve studies, as exhibited in the conformal gravity fits to the three representative galaxies shown below in Fig. (\ref{gangofthree}), the extracted $M/L$ values are typically as large as they could be without overshooting the rotation curve data in the inner galactic region. 

At each data point in each galaxy we can determine a value for the observed $g_{{OBS}}(R)=v_{{ OBS}}^2(R)/R$, and can thus construct the plot of $g(OBS)$ versus $g(NEW)$ for our 5791 points  that we show in Fig. (\ref{gobsgbar}). (Each point in the plot is constructed from the central value of each reported velocity and errors in velocity measurements are not incorporated.) 
We have applied the MLS formula to our data and given our different input parameters, find the mean fit shown in Fig. (\ref{gobsgbar}) with the slightly different $g_0=0.6\times 10^{-10}$ m s$^{-2}$ (the minimum is however quite shallow), but other than that we are in broad agreement with the analogous plot given in \cite{McGaugh2016}.\footnote{We have traced the difference between our fitted value of  $g_0=0.6\times 10^{-10}$ m s$^{-2}$ and the value MLS quote of $g_0=1.2\times 10^{-10}$ m s$^{-2}$ primarily to the fact that  MLS took a specific common $M/L$ value for all galaxies while we used luminous masses obtained from the conformal gravity fits themselves. (Moreover MLS themselves noted that their fits were not sensitive to the specific common value for $M/L$ that they used as such precision was not needed in order to establish a general trend.) The $M/L$ values that we used on average were twice the $M/L$ values MLS used, necessitating a reduction in two in our fitted value of $g_0$. In addition there was a non-negligible number  of cases in which  the NED/IPAC distances to galaxies  that we used differed significantly from the distances used by MLS (as quoted in the SPARC data basis). We should also note that for bright spirals the conformal gravity fitting leads to a luminous Newtonian expectation that  dominates the sharp initial rise of the inner region rotation curve (see e.g. the NGC 3198 plot that we provide in Fig. (\ref{gangofthree}). This rise is characteristic of the exponential disk formula given in (\ref{A2}), and this natural explanation of the inner region rise would be lost with smaller $M/L$ values, to then require some new dynamics in the inner regions of bright spirals.}

\begin{figure}[htpb!]
  \centering
    \includegraphics[width=1.5in,height=1.0in]{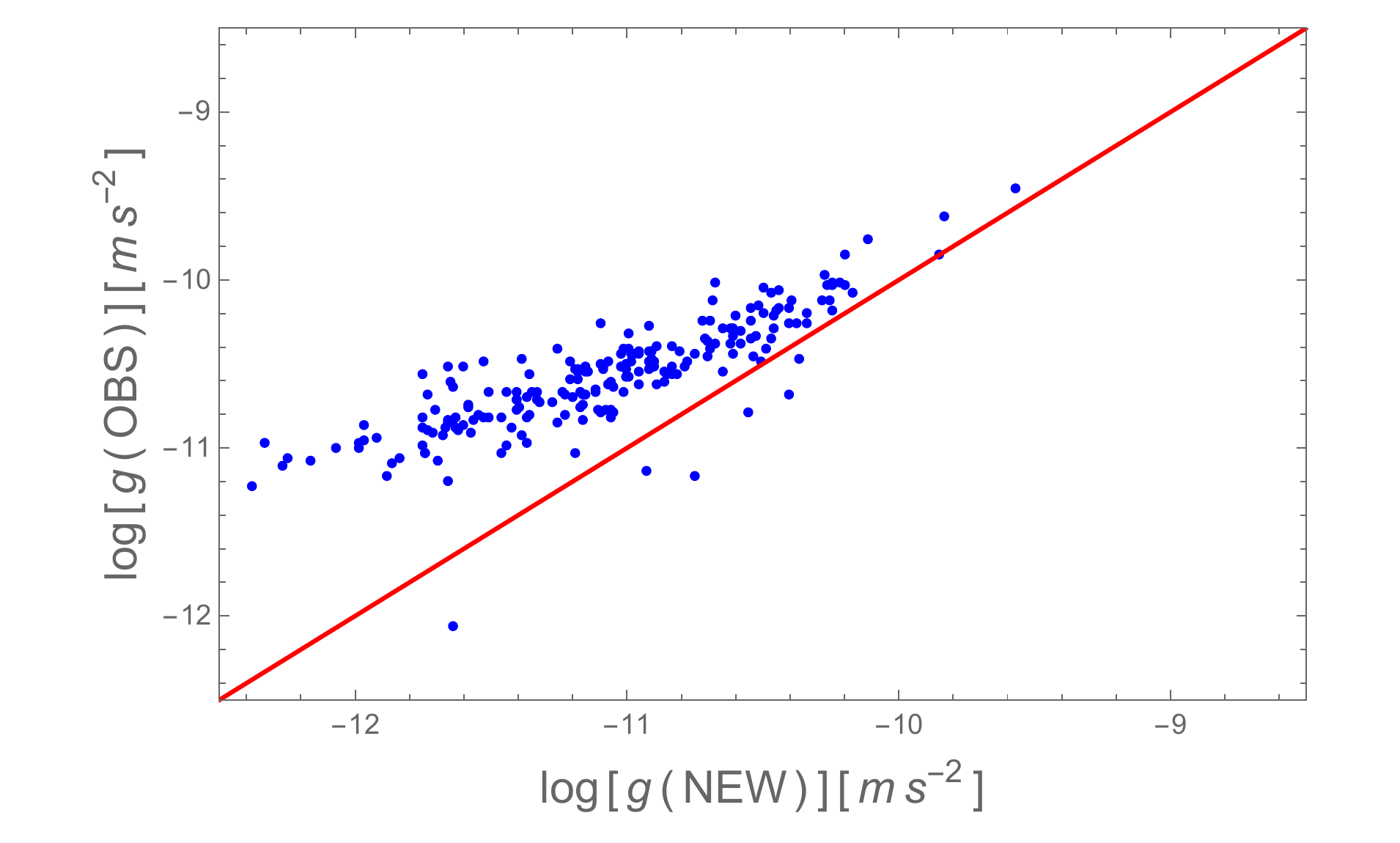}\qquad  \includegraphics[width=1.5in,height=1.0in]{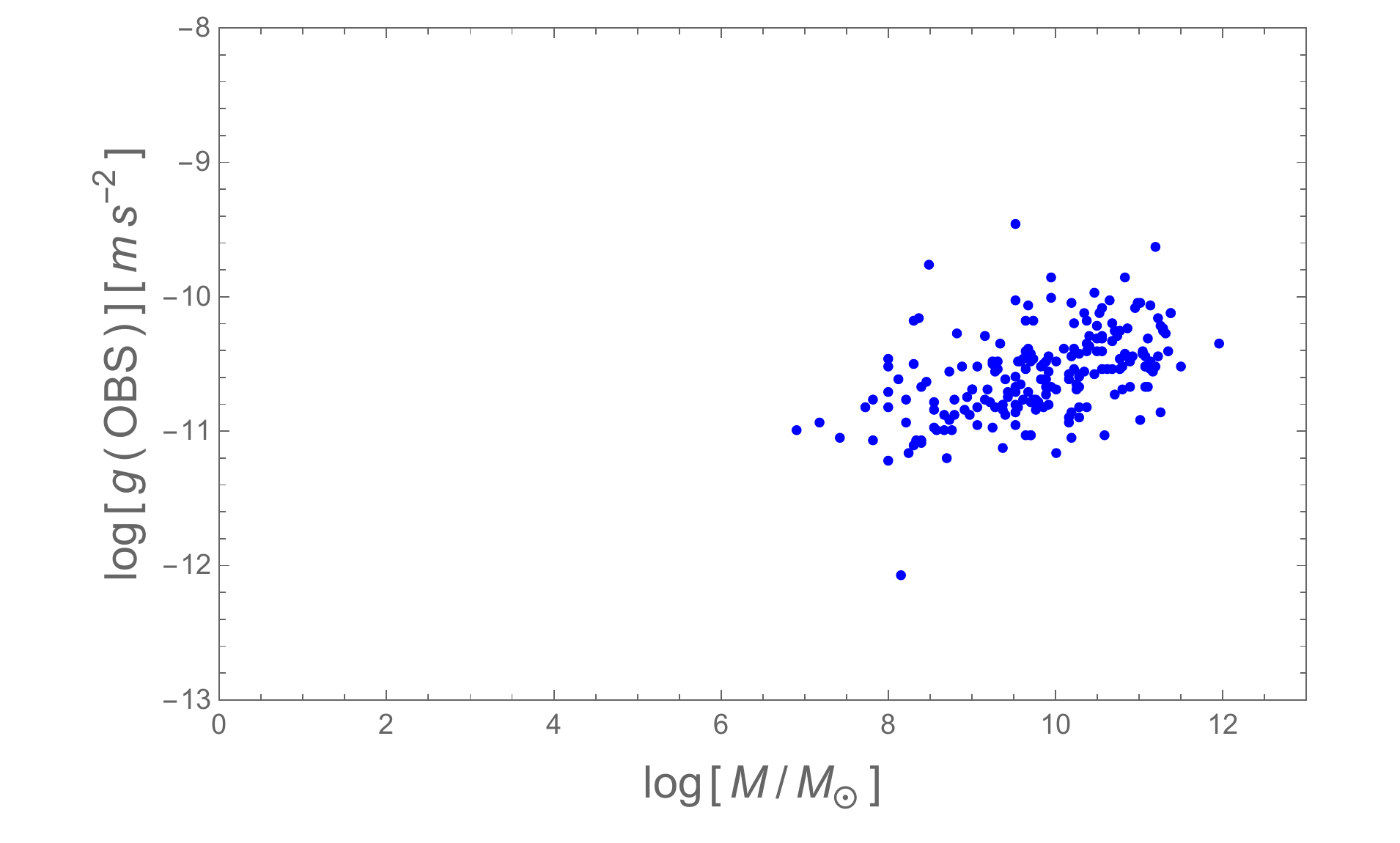}
  \caption{$g(OBS)$ versus $g(NEW)$ and $g(OBS)$ versus $M/M_{\odot}$ for the last data point in each galaxy.}
   \label{lastgbar}
\end{figure}

The realization that the band in Fig. (\ref{gobsgbar}) is quite narrow actually predates the work of \cite{McGaugh2016}, since in  the 141 galaxy study given in \cite{Mannheim2011} we had tabulated $g_{OBS}(R)$ at the last data points in each of the galaxies and found them to be remarkably close to each other in magnitude, with a value of order the quantity $\gamma_0c^2$ to be presented below. This same regularity is found to persist in the other 66 galaxies we have studied. For our 207 galaxy sample we illustrate this by plotting $g(OBS)$ versus $g(NEW)$ at each last data point in Fig. (\ref{lastgbar}). In Fig. (\ref{lastgbar}) we also plot  $g(OBS)$ versus the total mass, and this graph is particularly instructive since it essentially makes no reference to any gravity theory whatsoever, as it is basically a plot of the last centripetal acceleration versus luminosity galaxy by galaxy. This  regularity, together with the familiar Tully-Fisher relation that we discuss below, point in favor of the determining factor for rotation curve dynamics being, if anything, the luminous matter content rather than any possible non-luminous matter content in a galaxy, and any theory of galactic rotation curves would need to be able to account for it. 

\begin{figure}[htpb!]
  \centering
    \includegraphics[width=3.2in,height=1.0in]{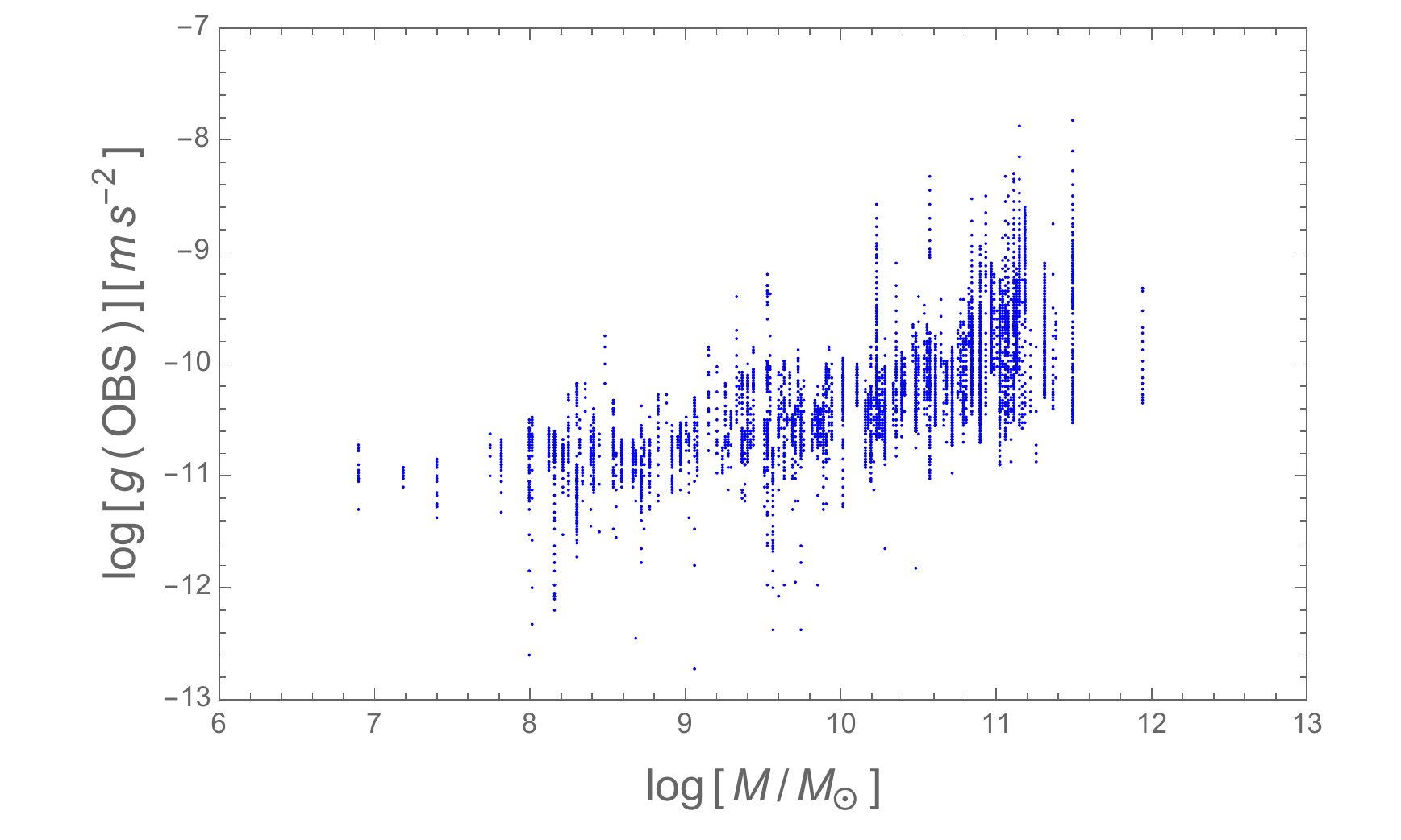}
  \caption{$g(OBS)$ versus $M/M_{\odot}$ for all points in each galaxy.}
   \label{gobsvmall}
\end{figure}
%
\begin{figure}[htpb!]
  \centering
    \includegraphics[width=3.2in,height=1.0in]{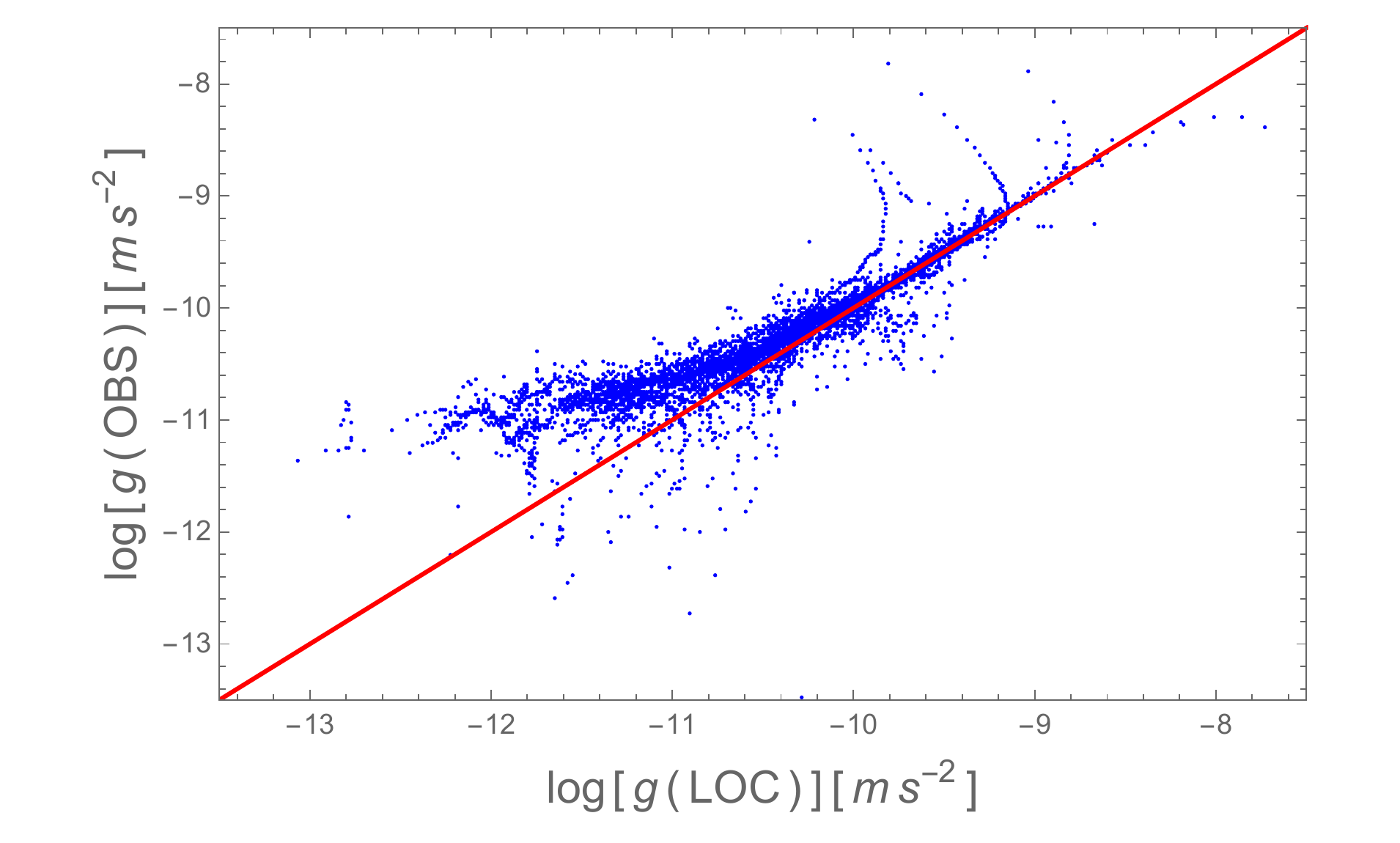}
      \caption{$g(OBS)$ versus the conformal gravity $g(LOC)$.}
   \label{gobsglum}
\end{figure}

To emphasize that the regularity is between $g(OBS)$ and luminous mass rather than between $g(OBS)$ and $g(NEW)$, in Figs. (\ref{gobsvmall}) and  (\ref{gobsglum}) we plot $g(OBS)$ versus the total visible galactic $M$ and $g(OBS)$ versus $g(LOC)$ for the entire 5791 points in our 207 galaxy sample, where $g(LOC)$ is the conformal gravity expectation due to all the local luminous matter in a galaxy as given in (\ref{A3}) below. With Figs. (\ref{gobsvmall}) and (\ref{gobsglum}) being very similar to Fig. (\ref{gobsgbar}), $g(OBS)$ has to be understood as being correlated with the luminous matter content in the galaxy per se rather than with its specific Newtonian expectation.  We will see below that while such an interpretation is valid, the same data also admit of an entirely different one.

\begin{figure}[htpb!]
\centering
  \includegraphics[width=1.1in,height=1.0in]{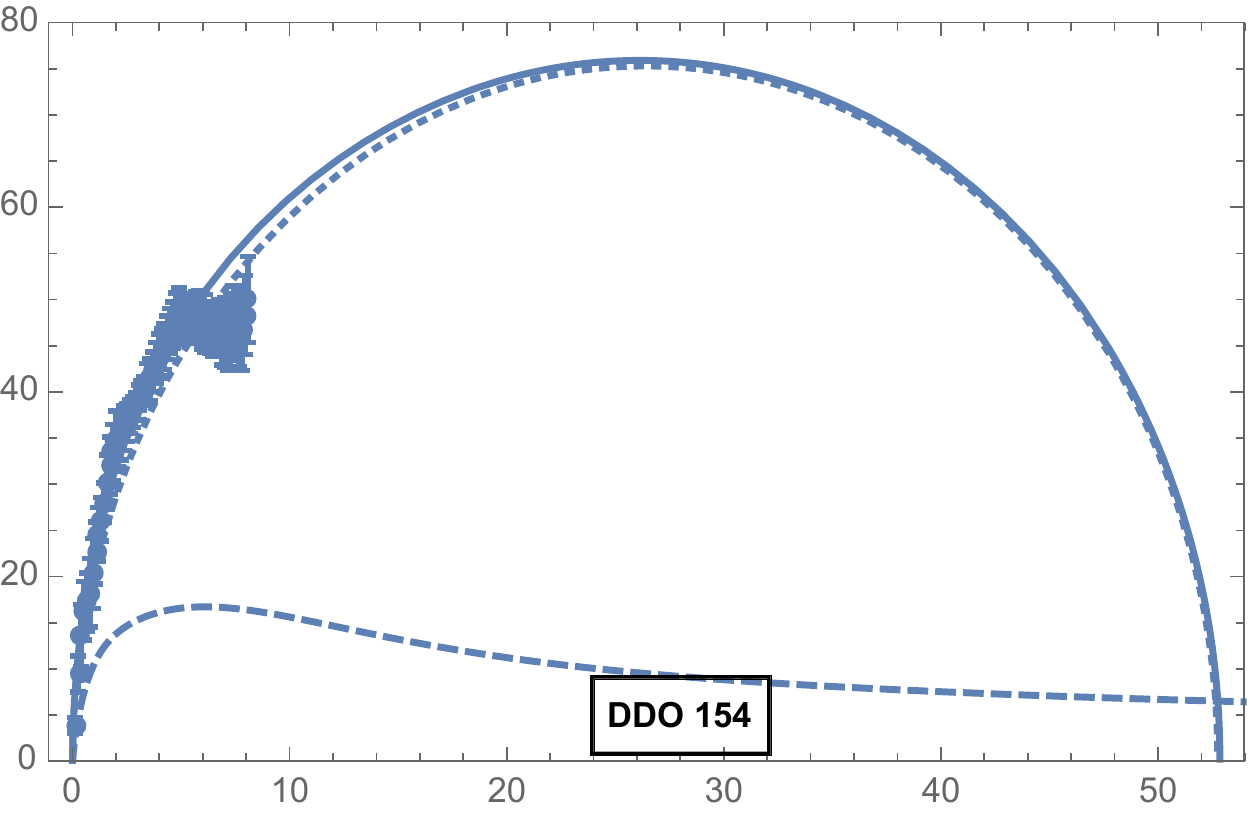}~
  \includegraphics[width=1.1in,height=1.0in]{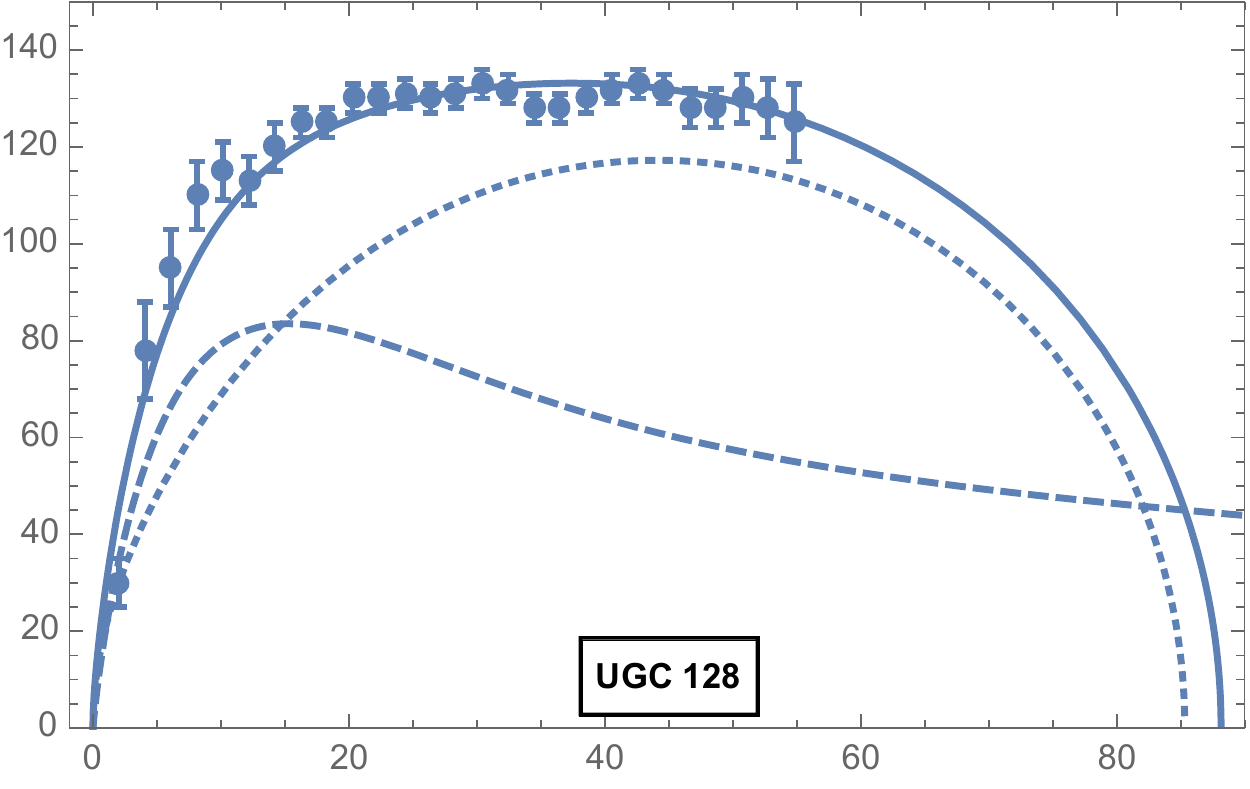}~
  \includegraphics[width=1.1in,height=1.0in]{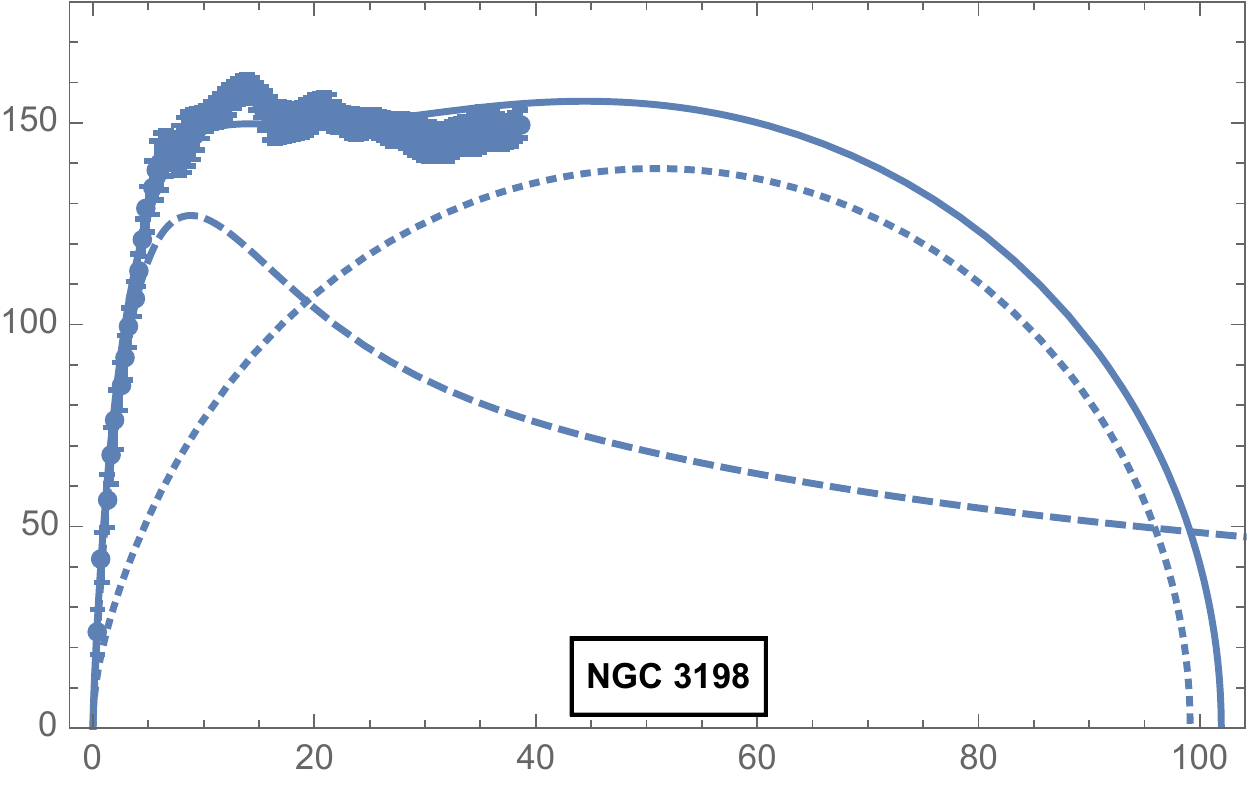}
\caption{Conformal gravity fits to some typical galactic rotation curves, with velocities in km~s$^{-1}$ and distances in kiloparsecs. The luminous Newtonian contribution is given by the dashed curve, the net contribution of the two linear potential terms and the quadratic potential term is given by the dotted curve, with the full curve giving the total contribution.}
\label{gangofthree}
\end{figure}

\section{Implications of the Data}

In our study of the 207 galaxies in our sample we have found that the conformal gravity theory provides very good point by point fitting to the 5791 point rotation curve data, with three examples (a dwarf DDO 154, a non-dwarf LSB UGC 128, and an HSB NGC 3198) being shown in Fig. (\ref{gangofthree}). Since Fig. (\ref{gobsgbar}) is based on the data points of the selfsame 207 galaxies, the conformal theory must also be compatible with Fig. (\ref{gobsgbar}). It is thus instructive to see how it achieves this not merely from the perspective of conformal gravity itself (to thus show that one can account for the systematics exhibited in Fig. (\ref{gobsgbar}) via a fundamental theory in which its velocity expectation (viz. (\ref{A4})) is derived from first principles), but from a more general perspective, as one can consider conformal gravity serving here as a foil. 

In the conformal gravity theory (viz. a pure metric theory of gravity that is based on the locally conformal invariant action $I_{ W}=-\alpha_g\int d^4x (-g)^{1/2}C_{\lambda \mu \nu \tau}C^{\lambda \mu \nu \tau}$ where $C_{\lambda \mu \nu \tau}$ is the conformal Weyl tensor) each star puts out a local ($LOC$) potential $V_{{LOC}}^{*}(r)=-\beta^{*}c^2/r+\gamma^{*} c^2 r/2$ \cite{Mannheim1989}, where $\gamma^*$ is a gravitational parameter associated with a linear potential. For an exponential  disk of stars the $V_{{LOC}}^{*}(r)$ potential generates a net  local term  
\begin{eqnarray}
g_{{  LOC}}(R)=g_{ NEW}(R)+(N^*\gamma^* c^2R/2R_0)I_1\left(x\right)
K_1\left(x\right)
\label{A3}
\end{eqnarray} 
due to the local visible material in the galaxy. In Newtonian gravity the force due to a spherically symmetric distribution of sources only depends on the sources that are interior to the point of observation since the solid angle grows like $r^2$ while the force falls like $1/r^2$. For any other potential there is no such exterior cancellation since the solid angle does not change as one changes the force. Thus for the conformal theory one has to to take the material exterior to any given galaxy (viz. the rest of the matter in the universe) into consideration. Since the matter exterior to a given galaxy does not depend on the galaxy of interest its effect is thus universal, with it being analytically found \cite{Mannheim1989,Mannheim2011} to lead to two global universal potential terms: a global linear term $\gamma_0c^2R/2$ due to the homogeneous cosmological background (which explains why there should be a universal acceleration in the first place), and a global quadratic potential term $-\kappa c^2R^2/2$ due to inhomogeneities in it such as clusters of galaxies. When taken together with the local terms the total centripetal acceleration given by the conformal theory (denoted by $CG$) is of the form \cite{Mannheim2011}
\begin{eqnarray}
g_{{  CG}}(R)=g_{{  LOC}}(R)
+\frac{\gamma_0c^2}{2}-\kappa c^2 R,
\label{A4}
\end{eqnarray} 
with large $R$ behavior 
\begin{eqnarray}
g_{{  CG}}(R)\rightarrow \frac{N^*\beta^*c^2}{R^2}+\frac{N^*\gamma^*c^2}{2}+\frac{\gamma_0c^2}{2}-\kappa c^2 R.
\label{A5}
\end{eqnarray} 

Very good point by point fitting to the entire 5791 data points in the 207 galaxy sample is obtained with only one free parameter per galaxy (viz. the visible $N^*$), with the $\gamma^*$, $\gamma_0$ and $\kappa$ parameters taking the fixed values $\gamma^*=5.42\times 10^{-41}~{\rm cm}^{-1}$, $\gamma_0=3.06\times 10^{-30}~{\rm cm}^{-1}$, $\kappa=9.54\times 10^{-54}~{\rm cm}^{-2}$
in every fit. The fitted values of these parameters show that $\gamma_0$ is indeed a cosmological scale,  that $\kappa$ is indeed a cluster of galaxies scale, and that dark matter is not needed for an understanding of the systematics of galactic rotation curves. When written in terms of an acceleration, we see that $\gamma_0c^2=2.76\times 10^{-11}$ m s$^{-2}$, a value that is characteristic of the values for $g(OBS)$ that are shown in the figures.
\begin{figure}[htpb!]
  \centering
    \includegraphics[width=3.2in,height=1.0in]{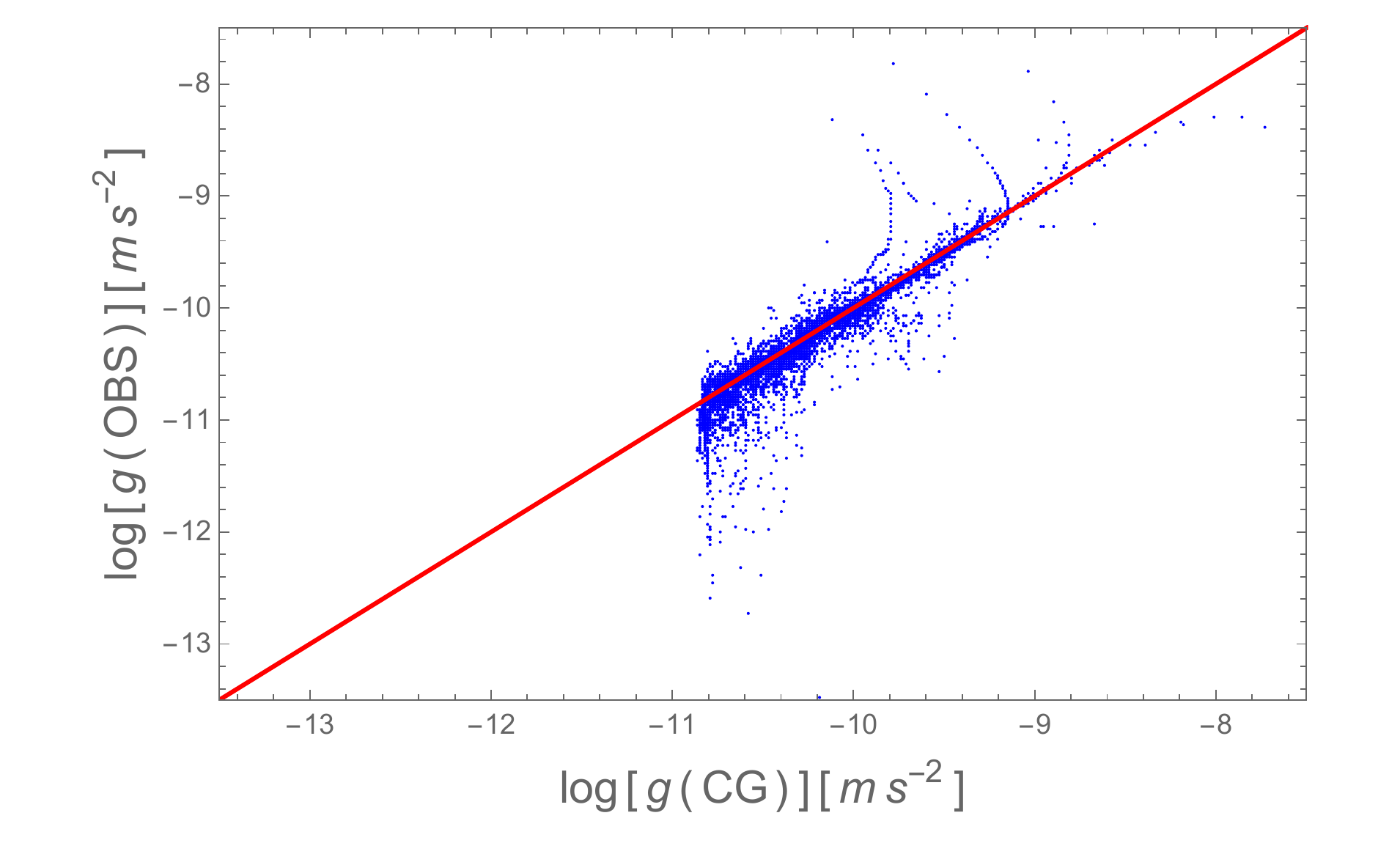}
  \caption{$g(OBS)$ versus the conformal gravity $g(CG)$. The solid diagonal is the line $g(OBS)=g(CG)$.}
   \label{cgfitall}
\end{figure}
%
\begin{figure}[htpb!]
  \centering
    \includegraphics[width=3.2in,height=1.0in]{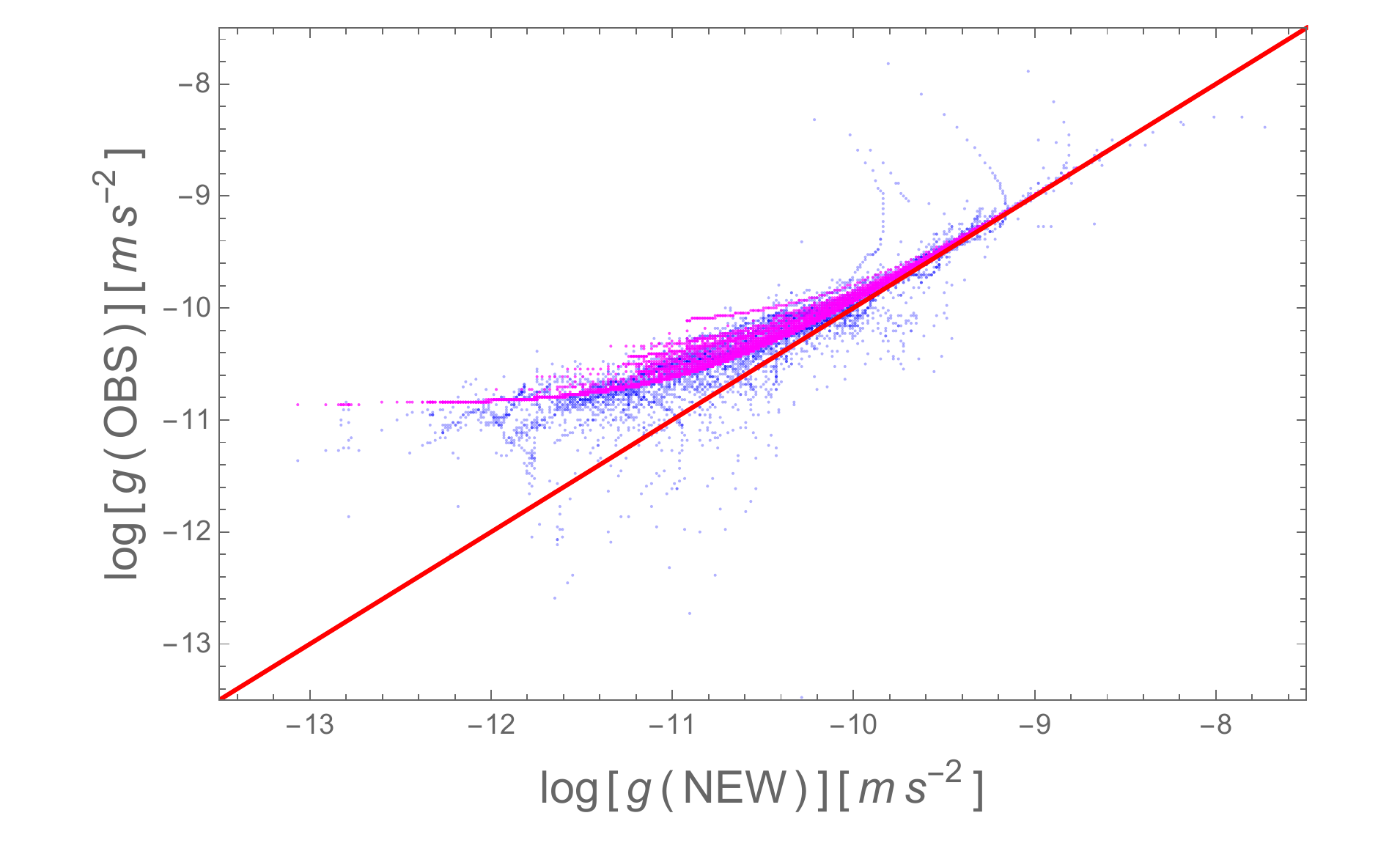}
  \caption{$g(CG)$ overlay of $g(OBS)$ versus $g(NEW)$. The solid lines other than the diagonal are the $g(CG)$ expectations. }
   \label{cgoverlayall}
\end{figure}

\begin{figure}[htpb!]
  \centering
    \includegraphics[width=3.2in,height=1.0in]{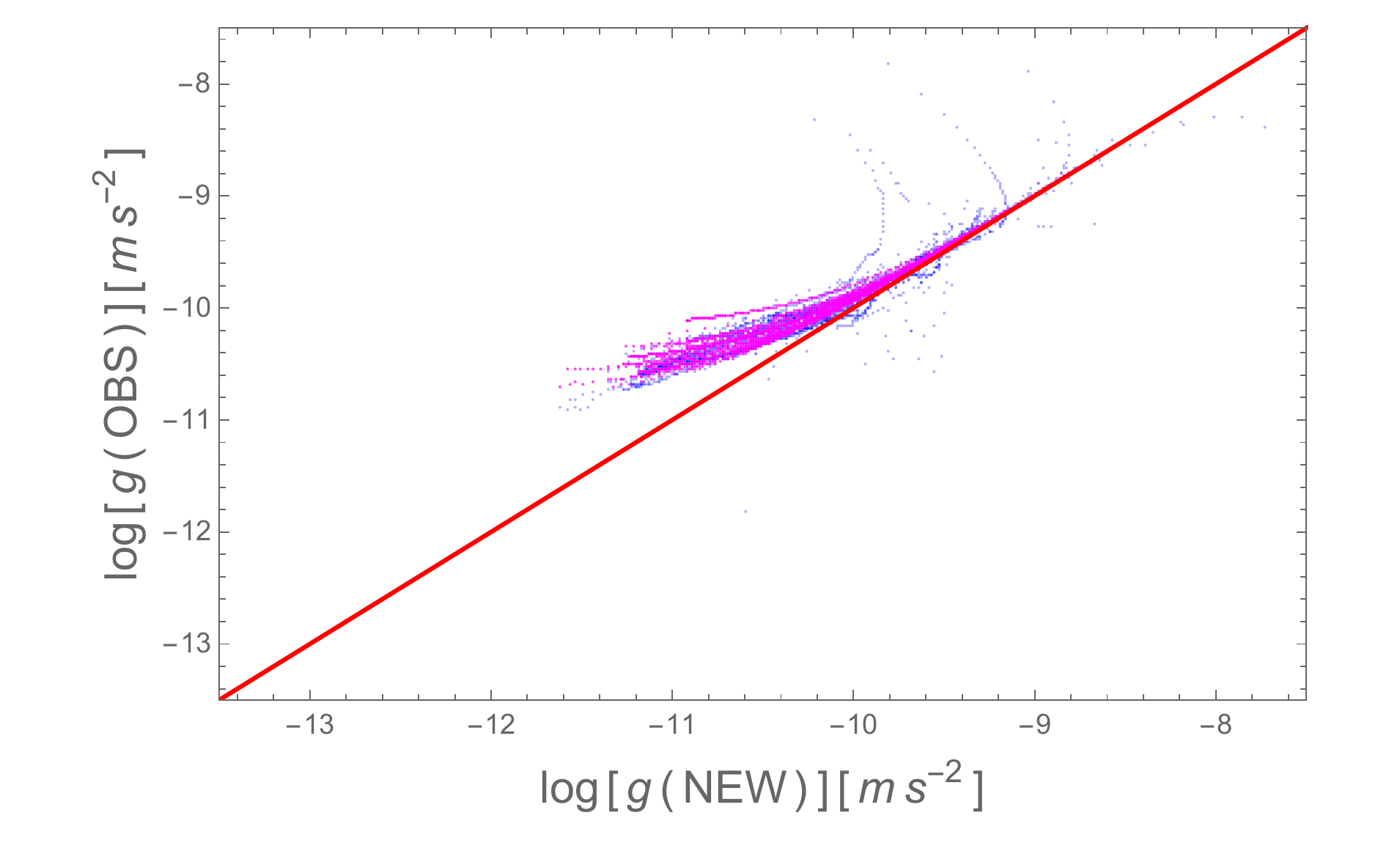}
  \caption{$g(CG)$ overlay of the HSB $g(OBS)$ versus  $g(NEW)$. The lines other than the diagonal are the $g(CG)$ expectations.}
   \label{cgoverlayhsb}
\end{figure}

\begin{figure}[htpb!]
  \centering
    \includegraphics[width=3.2in,height=1.0in]{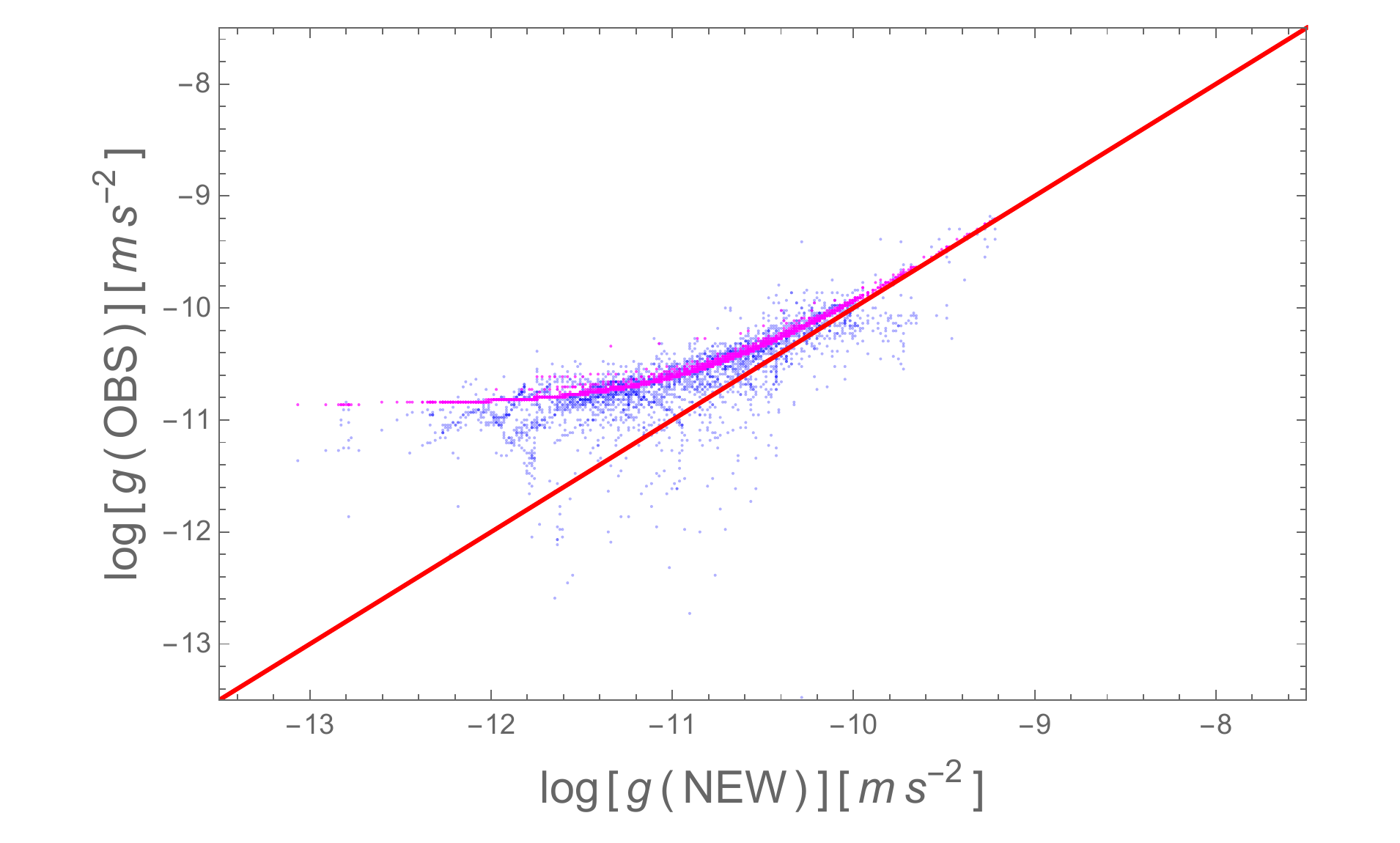}
  \caption{$g(CG)$ overlay of the LSB $g(OBS)$ versus $g(NEW)$. The lines other than the diagonal are the $g(CG)$ expectations.}
   \label{cgoverlaylsb}
\end{figure}

To demonstrate that the conformal gravity theory does fit all the data, in Fig. (\ref{cgfitall})  we have plotted $g(OBS)$ versus the conformal gravity $g(CG)$ of (\ref{A4}) for the entire 5791 data points.
Given the fit of Fig. (\ref{cgfitall}), we now overlay Fig. (\ref{gobsgbar}) with the conformal gravity $g(CG)$ predictions point by point, to obtain  Fig. (\ref{cgoverlayall}). As we see,  the conformal gravity predictions do not follow a single line but are spread out over the $g(OBS)$ band. To appreciate in what specific way the conformal gravity fits do cover the band, it is instructive to break $g(OBS)$ up into separate HSB and LSB pieces. This yields  Figs. (\ref{cgoverlayhsb}) and (\ref{cgoverlaylsb}).
%
%
\begin{figure}[htpb!]
  \centering
    \includegraphics[width=3.2in,height=1.0in]{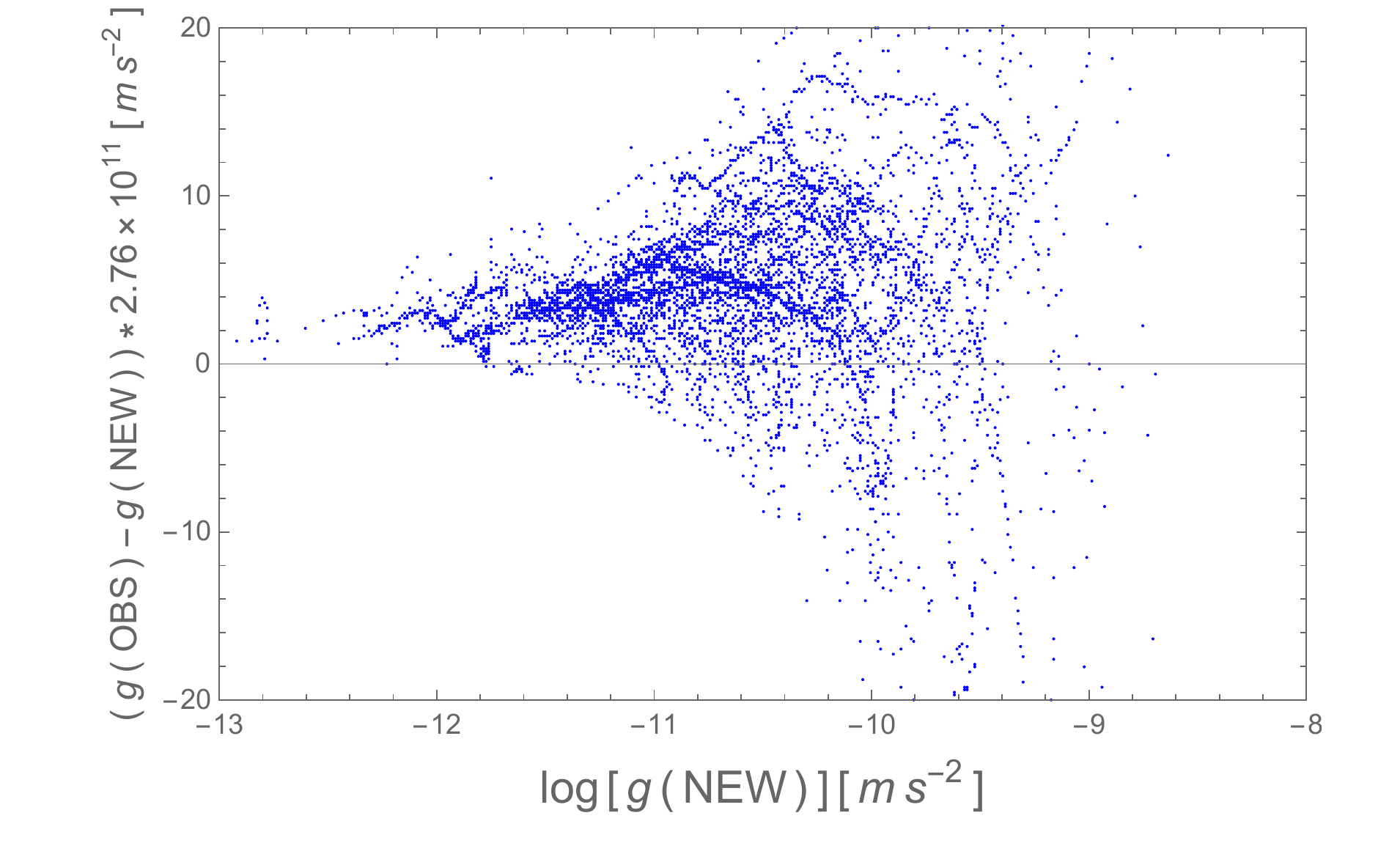}
  \caption{$g(OBS)-g(NEW)$ scaled by $\gamma_0c^2$ versus $g(NEW)$.}
   \label{asymptote}
\end{figure}

Figs. (\ref{cgoverlayhsb}) and (\ref{cgoverlaylsb}) exhibit a striking regularity, one that holds independent of conformal gravity per se: the data are essentially broken up into two distinct groups, the HSB galaxies all have large $g(OBS)$ values only, while the LSB galaxies extend to altogether smaller values. In addition, the conformal gravity HSB fits cover the width of the band, while the LSB fits center on a single curve. Now we had noted that the conformal theory fits the rotation curve data point by point. Thus the spread seen in the HSB Fig. (\ref{cgoverlayhsb}) does not represent scatter around a mean but is the actual data themselves. (With the $-\kappa c^2 R$ term in (\ref{A4}) only affecting the 20 or so largest galaxies in the sample \cite{Mannheim2011}, $g_{CG}(R)$ is otherwise bounded from below by its $N^*=0$ value, to thus give a band.) The width in Fig. (\ref{gobsgbar}) represents bona fide physical data, and even if it were, as suggested in \cite{McGaugh2016},  to be attributed solely to scatter around a mean,\footnote{For the band in Fig. (\ref{gobsgbar}) to be due solely to observational uncertainties, it would have to be the case that for each specific galaxy one would have to be able to adjust galactic input parameters within observational bounds so that all the rotation curve points in that galaxy would then simultaneously agree with a formula such as $g(OBS)=g(MLS)$. And then, one would have to be able to do this for the entire 207 galaxy sample, galaxy by galaxy. Also we would note that the band only appears in the HSB sample, and here uncertainties are much smaller than in the LSB sample. In addition, we should note that fitting Fig. (\ref{gobsgbar}) is not as stringent as actually fitting individual rotation curves themselves point by point, since a error of order $\delta$ in velocity translates into an error of order $2\delta$ in $g(OBS)$.} even then one could say exactly the same of Fig. (\ref{cgfitall}), with the conformal gravity interpretation of Fig. (\ref{gobsgbar}) still being fully justifiable.

Fitting for the LSB sample is quite different. Here as one reduces $N^*$ one reduces $g(NEW)$. Then, until one is at an $R$ large enough for the $-\kappa c^2R$ term in (\ref{A4}) to be of consequence, the dominant term in $g_{CG}(R)$ as $N^*$ reduces becomes $\gamma_0c^2/2$. This term is universal and galaxy independent, and thus one single curve dominates the small $g(NEW)$ region of the LSB sample, and leads  as $g(NEW)\rightarrow 0$ to a constant asymptote, as shown in Figs. (\ref{cgoverlaylsb}) and (\ref{asymptote}), the latter of which plots a scaled up $g(OBS)-g(NEW)$ in the central region of the data points. Thus as $g(NEW)\rightarrow 0$,  $g(OBS)$ becomes independent of the luminous matter content altogether. This particular behavior is exhibited in the fit to the dwarf DDO 154 as $N^*$ is so low that the rotation curve is dominated by $\gamma_0$  not just in the outer region but in the inner region too. Dwarfs are particularly interesting for rotation curve studies since the luminous Newtonian shortfall is evident even in the inner region, with these galaxies immediately revealing the full extent of the galactic missing mass problem.

The behavior of the conformal  theory as $g(CG)\rightarrow 0$ differs substantially from that associated with the $g(MLS)$ function, as $g(MLS)$ asymptotes to $g(NEW)^{1/2}$ times a constant, to thus never become independent of the luminous Newtonian contribution. While a possible discrimination between these various options awaits more small $g(NEW)$ data,\footnote{A first step in this direction has recently been taken by  Lelli, McGaugh, Schombert, and Pawloski in a follow up paper \cite{Lelli2017}, which  showed that the data might indeed be becoming independent of $g(NEW)$ at small $g(NEW)$. Specifically, they augmented the spiral galaxy data with some dwarf spheroidal data and some late type galaxy data, which showed such flattening off at very low $g(NEW)$. Lelli, McGaugh, Schombert, and Pawloski even considered changing the $g(MLS)$ formula because of this flattening off (by adding on to $g(MLS)$ a term $\hat {g}\exp(-(g(NEW)g_0/\hat{g}^2)^{1/2})$ where $\hat{g}$ is a new free parameter, and in their paper characterized the data as exhibiting a possible "acceleration floor". As we have seen, the existence of such an acceleration floor follows naturally in the conformal gravity theory.} our analysis here does show that  from Fig. (\ref{gobsgbar}) one cannot infer that $g(OBS)$ will always depend on $g(NEW)$. 

At very large $R$ the effect of the $\gamma_0c^2/2$ term in (\ref{A4}) is overcome by the $-\kappa c^2 R$ term, causing rotation velocities to start to drop and eventually come to zero, just as anticipated in Fig. (\ref{gangofthree}). Since $v^2$ cannot go negative, in the conformal theory galaxies have to have a maximum size, a size that is fixed via an interplay between the local and global terms in (\ref{A4}). Since such an outcome is not to be expected in theories in which rotational velocities are asymptotically flat, through the large $R$ behavior of rotation curves one can test the conformal theory, and also one can explore the $g(NEW)\rightarrow 0$ limit of Fig. (\ref{gobsgbar}) when $N^*$ is large. Since the conformal gravity theory does involve both local and global effects but no dark matter, postulating the presence of dark matter within galaxies can be viewed as being nothing more than an attempt to describe global physics in purely local terms. 

In regard to dark matter, we also note that  the challenge to it can be summarized by stating that there is universality in the data that is for the moment not accounted for by dark matter theories. Various alternate theories such as MOND \cite{Milgrom1983a}, conformal gravity, and Moffat's Modified Gravity Theory (MOG) \cite{Moffat2013} that fit rotation curves without any dark matter whatsoever, all do so with universal parameters, with only the $M/L$ ratio varying from one galaxy to the next. In contrast, current dark matter models possess no such universality, and with each dark matter halo having at least two free parameters, to fit the 207 galaxy Fig. (\ref{gobsgbar}) dark matter models need 414 more free parameters than the three alternate theories.

\section{Distance-Dependent Regularity}

\begin{figure}[htpb]
  \centering
    \includegraphics[width=3.2in,height=1.0in]{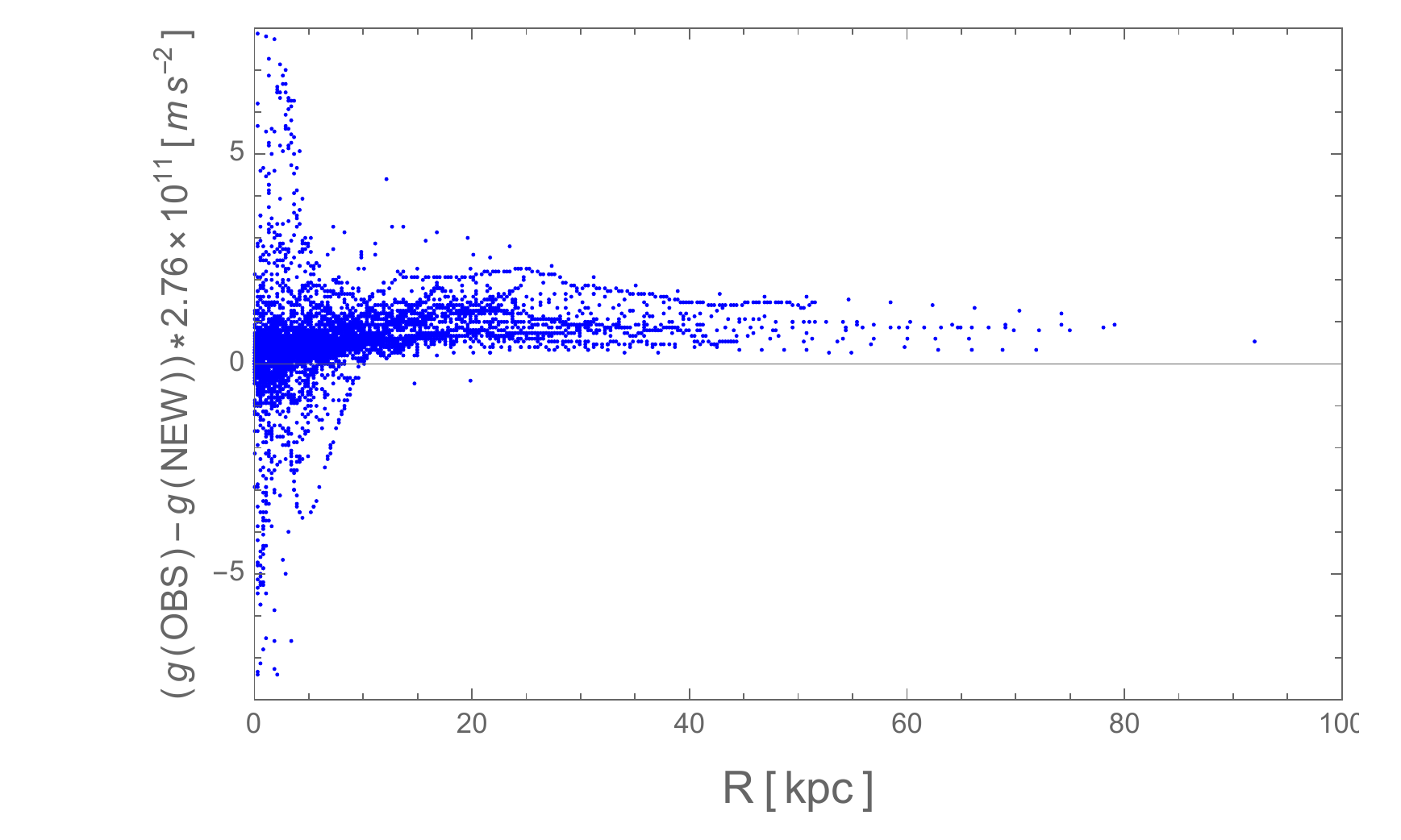}
  \caption{$g(OBS)-g(NEW)$ scaled by $\gamma_0c^2$ versus $R$.}
   \label{versusr}
\end{figure}
Even though the $g(OBS)$ versus $g(NEW)$ plot does exhibit a remarkable structure, for practical applications it has the drawback that it only involves $v^2/R$ type ratios. Thus one can obtain a given value of $g(NEW)$ for many  values of $v$ by choosing differing values for $R$, with this being equally true of any value of $g(OBS)$ that might be read off from Fig. (\ref{gobsgbar}) using this given value of $g(NEW)$. It would thus be instructive to have a universal plot for all galaxies that does depend on $R$. To thus end rather than plot $g(OBS)-g(NEW)$ as function of $g(NEW)$ (as in Fig. (\ref{asymptote})), we have found it instructive to instead plot $g(OBS)-g(NEW)$ as a function of $R$ point by point for the 5791 points in our sample. This yields Fig. (\ref{versusr}).

Inspection of Fig. (\ref{versusr}) shows that by 10 kpc there is a luminous Newtonian shortfall at every single data point in our sample, with $g(OBS)-g(NEW)$ being positive for all such points. Moreover, the actual amount of the shortfall above 10 kpc is confined to a very narrow horizontal band, to thus be independent of $R$. Given the asymptotic behavior of $g_{CG}(R)$ exhibited in (\ref{A5}), we see that apart from the $-\kappa c^2R$ term (a term that is only important for the 20 or so very largest of the galaxies), the $N^*\gamma^*c^2/2$ and $\gamma_0c^2/2$ contributions are in fact constant, with the sum taking the value $\gamma_0c^2$ when $N^*=\gamma^0/\gamma^*=5.65\times10^{ 10}$, viz. a mass value typical of bright 
spirals.\footnote{As shown in the 141 galaxy conformal gravity fits given  in \cite{Mannheim2011}, within reported errors in the measured velocities the asymptotic (\ref{A5}) gives a very good accounting of the  large $R$ data.} Independent of its potential relevance to conformal gravity, Fig. (\ref{versusr}) is of general interest because it encapsulates large $R$ departures from the luminous Newtonian expectation in a very compact and direct way.

\section{First Principles Derivation of the Tully-Fisher Regularity}

\begin{figure}[htpb]
  \centering
    \includegraphics[width=3.2in,height=1.0in]{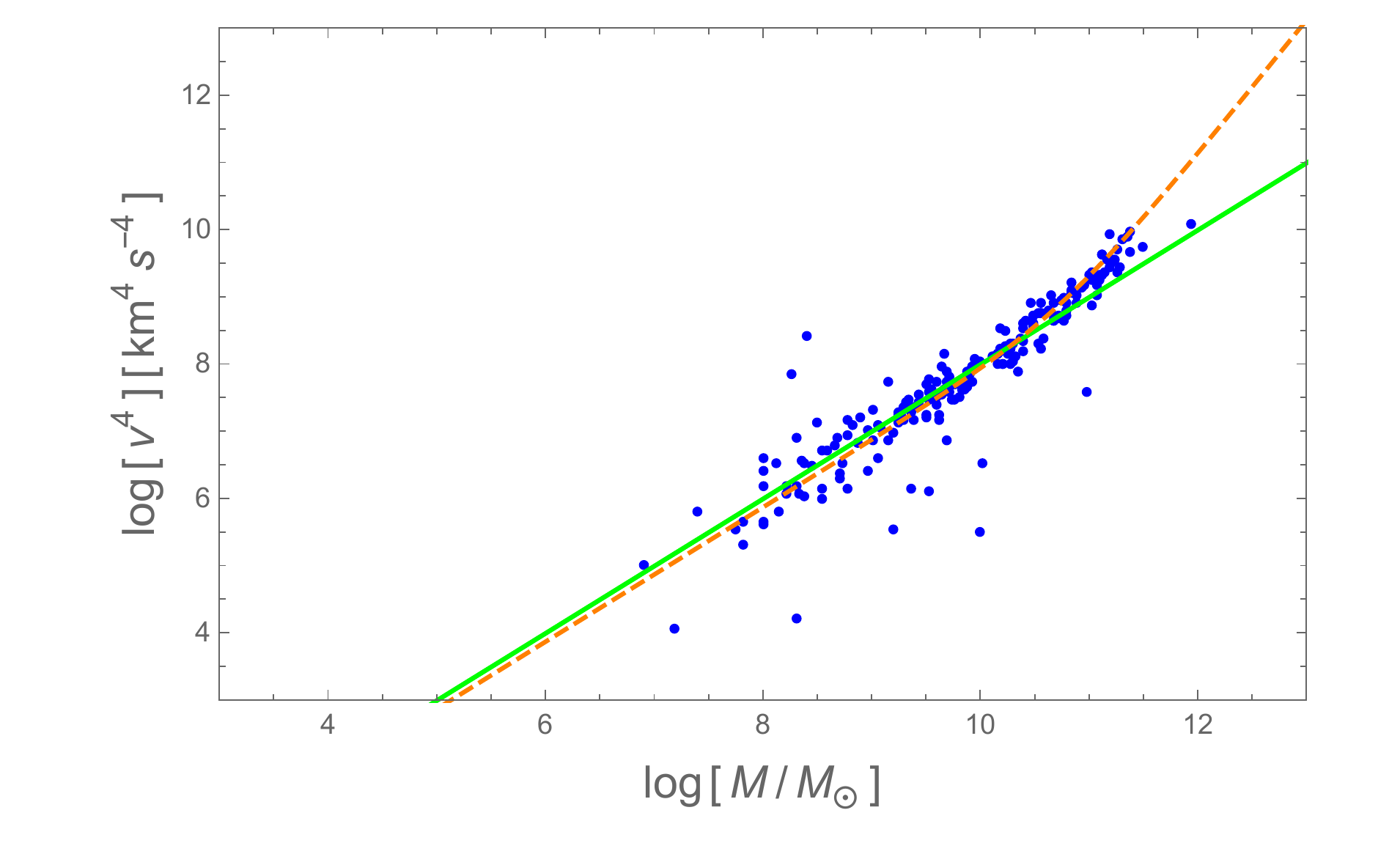}
  \caption{$v^4$ versus $M$ for the last data point point in each of the 207 galaxies. Overlaid are $v^4=AM/M_{\odot}$ (continuous curve) and $v^4=B(M/M_{\odot})(1+N^*/D)$ (dashed curve).}
   \label{tullyfisher}
\end{figure}
%

\begin{figure}[htpb]
  \centering
    \includegraphics[width=3.2in,height=1.0in]{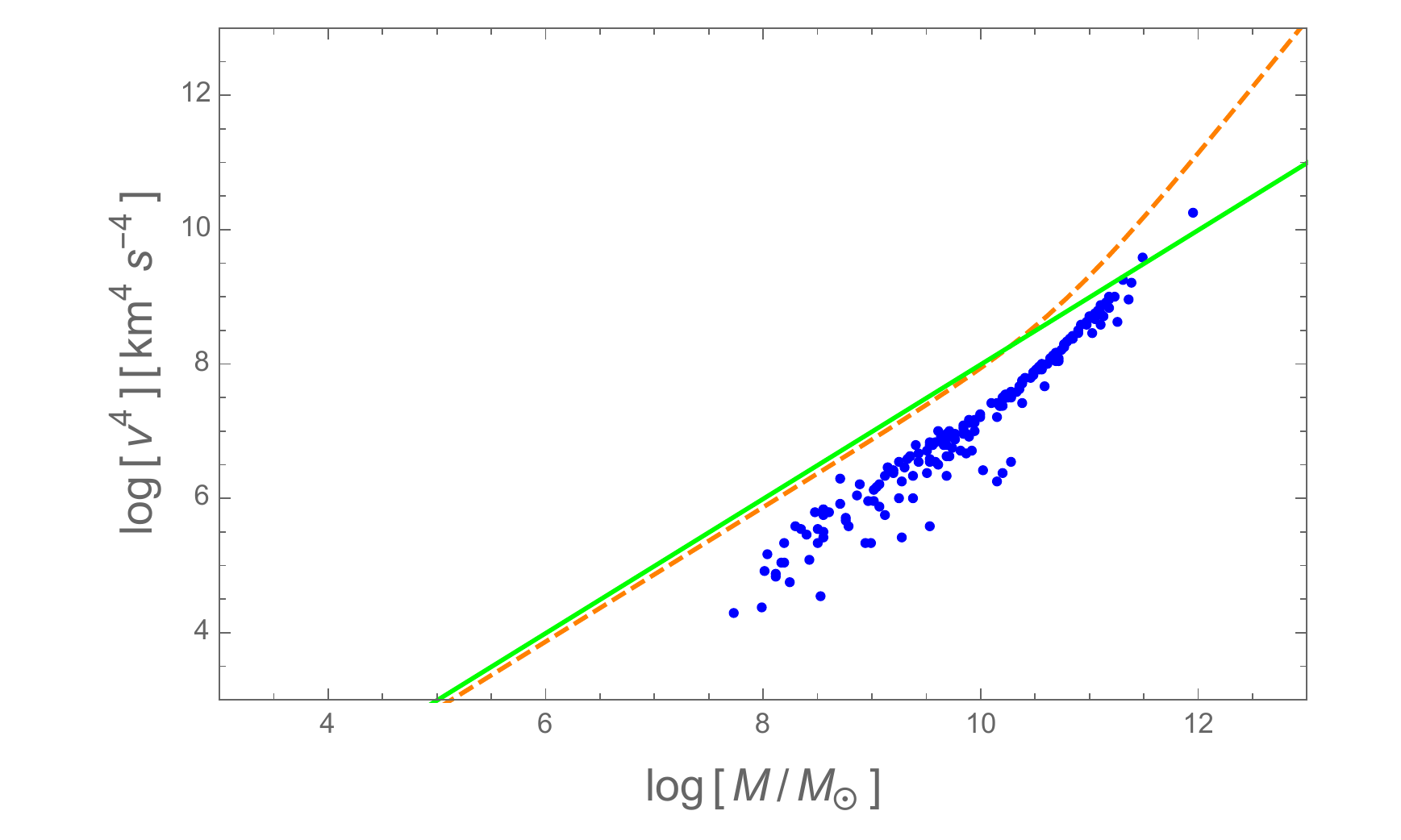}
  \caption{$v^4$ versus $M$ for the crossover point in each of the 207 galaxies. Overlaid are the $v^4=AM/M_{\odot}$ (continuous curve) and $v^4=B(M/M_{\odot})(1+N^*/D)$ (dashed curve) given in Fig. (\ref{tullyfisher}) for the last data points.}
\label{crossovertf}
\end{figure}
For many spiral galaxies it has been found phenomenologically that the average rotational velocity obeys the Tully-Fisher relation $v^4\sim L$. Thus up to mass to light ratios we can set $v^4=AM/M_{\odot}$ in a convenient normalization. While the Tully-Fisher relation is usually stated in terms of the stellar luminosity, the more relevant quantity for velocities is the total mass $M$, which here therefore includes not just the stellar mass (disk and bulge) but the gas mass as well. (McGaugh et. al. \cite{McGaugh2000} have referred to the use of the total $M$ rather than $L$ as the baryonic Tully-Fisher relation.) With the velocity at the last data point in each galaxy typically being representative of the average velocity in each galaxy (rotation curves being close to flat), a plot of $v^4=AM/M_{\odot}$ for the last data point in each of our 207 sample galaxies is provided as the continuous curve in Fig. (\ref{tullyfisher}), a fit that gives an extracted value of $A=0.0098$ km$^4$s$^{-4}$. As can be seen in the fits to UGC 128 and NGC 3198, in conformal gravity fits to rotation curves the conformal gravity contribution and the Newtonian contribution typically cross in a region far enough out from the center of the galaxy so that the galaxy can be treated as a point source, and close enough in that the quadratic term in (\ref{A4}) is negligible. Thus at that point one can set $v^2=\beta^*c^2N^*/R+(\gamma^*N^*+\gamma_0)c^2R/2$ and $\beta^*c^2N^*/R=(\gamma^*N^*+\gamma_0)c^2R/2$ for an $R$ that depends on each galaxy, and thus at that point one can set $v^4=B(M/M_{\odot})(1+N^*/D)$ where $B=2c^2M_{\odot}G\gamma_0=0.0074$ km$^4$s$^{-4}$ and $D=\gamma_0/\gamma^*=5.65\times10^{ 10}$. Since the velocities at the last data points do not differ much from those at the crossover points in each galaxy, at the last data points we plot $v^4=B(M/M_{\odot})(1+N^*/D)$ as the dashed curve in Fig. (\ref{tullyfisher}). (To show that there is little difference beween the velocities at the last data points and those at the crossover points, in Fig. (\ref{crossovertf}) we also plot $v^4$ versus $M$ at the crossover points.) Since very few galaxies have $N^*>5.65\times10^{10}$, conformal gravity effectively leads to  $BM/M_{\odot}<v^4<2BM/M_{\odot}$, to not just be in agreement with  Fig. (\ref{tullyfisher}), but to also provide a first principles derivation of the  Tully-Fisher relation.

\section{Summary}

To summarize, we note that the conformal gravity fitting of the HSB sample given in Fig. (\ref{cgoverlayhsb}) suggests that the width seen in Fig. (\ref{gobsgbar}) is physical and not just scatter. The very low $g(NEW)$ limit of the fitting of the LSB sample given in Fig. (\ref{cgoverlaylsb}) suggests that at small $g(NEW)$ the quantity $g(OBS)$ is limiting to a value that is independent of $g(NEW)$, and that $g(OBS)$ is not necessarily determined by $g(NEW)$ alone. The plot of $g(OBS)-g(NEW)$ against $R$ shows that $g(OBS)-g(NEW)$ is asymptoting to a value that is independent of $R$, just as expected in the conformal theory. Also, the conformal theory provides a first principles derivation of the Tully-Fisher relation. The plots presented in this paper point to regularities in the data that need to be accounted for in any theory of rotation curves. Now none of this is to say that conformal gravity is necessarily to be preferred over any other theory. Nonetheless, from the perspective of conformal gravity our fits indeed show that $g(OBS)$ is fixed by luminous matter alone, but the luminous matter that is relevant is not just from within galaxies but from the rest of the visible universe as well.

The authors wish to thank Dr. S. S. McGaugh, Dr. F. Lelli, and Dr. J. M. Schombert for providing them with the SPARC database, and wish to thank S. Chaykov, Dr. B.  Placek, and Dr. F.  Rueckert for discussions.  This research has made use of the NASA/IPAC Extragalactic Database (NED) which is operated by the Jet Propulsion Laboratory, California Institute of Technology, under contract with the National Aeronautics and Space Administration.

\end{document}